\documentclass[namedreferences]{solarphysics}

\usepackage[hyperref,optionalrh]{spr-sola-addons} 
\usepackage{graphicx}        
\usepackage{color}           
\usepackage{breakurl}        
\usepackage{bm}

\newcommand{\aap}{    {\it Astron. Astrophys.}}

\newcommand{\apj}{    {\it Astrophys. J.}}
\newcommand{\apjl}{   {\it Astrophys. J. Lett.}}
\newcommand{\apjs}{   {\it Astrophys. J. Sup.}}
\newcommand{\apss}{   {\it Astrophys. Space Sci.}}

\newcommand{\grl}{    {\it Geophys. Res. Lett.}}

\newcommand{\mnras}{  {\it Mon. Not. Roy. Astron. Soc.}}
\newcommand{\nat}{    {\it Nature}}
\newcommand{\na}{    {\it New. Astron.}}

\newcommand{\pasp}{   {\it Pub. Astron. Soc. Pac.}}
\newcommand{\pasj}{   {\it Pub. Astron. Soc. Japan}}

\newcommand{\solphys}{{\it Solar Phys.}}
 
\newcommand{\ssr}{    {\it Space Sci. Rev.}} 
\newcommand{\speed}[1]{#1 km~s${}^{-1}$}
\newcommand{\acc}[1]{#1 km~s${}^{-2}$}

\chardef\us=`\_

\begin{document}
\begin{article}
\begin{opening}

\title{CME-Driven and Flare-Ignited Fast Magnetosonic Waves Successively Detected in a Solar Eruption}

\author[addressref={aff1,aff2,aff3,aff4}]{\inits{X.P.}\fnm{Xinping}~\lnm{Zhou}\orcid{0000-0001-9374-4380}} 
\author[addressref={aff1,aff2,aff3,aff4},corref,email={ydshen@ynao.ac.cn}]{\inits{Y.D.}\fnm{Yuandeng}~\lnm{Shen}\orcid{0000-0001-9493-4418}}
\author[addressref={aff3,aff4}]{\inits{J.T.}\fnm{Jiangtao}~\lnm{Su}}
\author[addressref={aff1,aff4}]{\inits{Z.H.}\fnm{Zehao}~\lnm{Tang}\orcid{0000-0003-0880-9616}}
\author[addressref={aff1,aff4}]{\inits{C.R.}\fnm{Chengrui}~\lnm{Zhou}}
\author[addressref={aff1,aff4}]{\inits{Y.D.}\fnm{Yadan}~\lnm{Duan}\orcid{0000-0001-9491-699X}}
\author[addressref={aff1,aff4}]{\inits{S.}\fnm{Song}~\lnm{Tan}}

\address[id=aff1]{Yunnan Observatories, Chinese Academy of Sciences, Kunming 650011, China}
\address[id=aff2]{State Key Laboratory of Space Weather, Chinese Academy of Sciences, Beijing 100190, China}
\address[id=aff3]{Key Laboratory of Solar Activity, National Astronomical Observatories, Chinese Academy of Sciences, Beijing 100012, China}
\address[id=aff4]{University of Chinese Academy of Sciences, Beijing 10049,China}
\runningauthor{Zhou et al.}
\runningtitle{Simultaneous Detection of CME-driven and Flare-ignited Waves in a Flare/CME Event}

\begin{abstract}
We present {\em SDO}/AIA observation of three types of fast-mode propagating magnetosonic waves in a {\em GOES} C3.0 flare on 2013 April 23, which was accompanied by a prominence eruption and a broad coronal mass ejection (CME). During the fast rising phase of the prominence, a large-scale dome-shaped extreme ultraviolet (EUV) wave firstly formed ahead of the CME bubble and propagated at a speed of about \speed{430} in the CME's lateral direction. One can identify the separation process of the EUV wave from the CME bubble. The reflection effect of the on-disk counterpart of this EUV wave was also observed when it interacted with a remote active region. Six minutes after the first appearance of the EUV wave, a large-scale quasi-periodic EUV train with a period of about 120 seconds appeared inside the CME bubble, which emanated from the flare epicenter and propagated outward at an average speed up to \speed{1100}. In addition, another narrow quasi-periodic EUV wave train was observed along a closed-loop system connecting two adjacent active regions, which also emanated from the flare epicenter, propagated at a speed of about \speed{475} and with a period of about 110 seconds. We propose that all the observed waves are fast-mode magnetosonic waves, in which the large-scale dome-shaped EUV wave ahead of the CME bubble was driven by the expansion of the CME bubble, while the large-scale quasi-periodic EUV train within the CME bubble and the narrow quasi-periodic EUV wave train along the closed-loop system were excited by the intermittent energy-releasing process in the flare. Coronal seismology application and energy carried by the waves are also estimated based on the measured wave parameters.
\end{abstract}
\end{opening}

\section{Introduction}
Large-scale magnetohydrodynamic (MHD) waves in the solar atmosphere such as umbra and penumbra of sunspots running waves\citep[e.g.,][]{2017NewA...51...86Z,2017Ap&SS.362...46Z}, chromospheric Moreton waves, and coronal extreme ultraviolet (EUV) waves have been investigated for many years  \citep[e.g.,][]{1960PASP...72..357M,2012ApJ...752L..23S,2018MNRAS.474..770K,2020ApJ...894...30W}. The measured physical parameters of MHD waves are crucial for diagnosing their physical nature, excitation mechanism, and energy transport in the solar atmosphere. Large-scale EUV disturbances are typically interpreted as fast-mode magnetosonic waves, which can propagate globally and generally interact with various coronal structures. For example, they can cause the oscillation of remote filaments and coronal loops \citep{2002SoPh..206...69S,1999SoPh..190..467W,1999ApJ...520..880A,2009SSRv..149..283T,2013ApJ...773..166L,2014ApJ...786..151S,2013SoPh..282..523K,2019ApJ...873...22S}. The wave nature can be evidenced through reflection, refraction, and transmission effects when EUV waves interacted with remote coronal holes \citep{2009ApJ...691L.123G,2012ApJ...746...13L,2012ApJ...756..143O,2018ApJ...864L..24L,2019ApJ...878..106H,2019ApJ...870...15L}, active regions (\cite{1999SoPh..190..467W},\cite{2002ApJ...574..440O},\cite{2007ApJ...655.1134O},\cite{2013SoPh..288..255K},\cite{2013ApJ...773L..33S},\cite{2013A&A...553A.109K}). Especially, the oscillation phenomena of the filaments or prominences after impingement are very suitable for local seismology, which is a good technique for estimating the magnetic field strength of the filament and the surrounding coronal environment. EUV waves also carry critical information that can be used to diagnose some important physical parameters of the corona such as the magnetic field strength, which is an area yet to be fully exploited (e.g., \cite{2011ApJ...741L..21L,2019ApJ...873...22S,1970PASJ...22..341U,2005LRSP....2....3N,2017Ap&SS.362...46Z,2021ApJ...908L..37M}).

Large-scale coronal EUV waves can propagate across a large fraction the solar disk with a typical speed of \speed{200--400} \citep{2009ApJS..183..225T}, which were firstly observed by the EUV Imaging Telescope (EIT; \cite{1995SoPh..162..291D}) onboard the {\em Solar and Heliospheric Observatory spacecraft} ({\em SOHO}; \citep{1997SoPh..175..571M,1998GeoRL..25.2465T}). Over the past two decades, though a large number of wave events have been analyzed with various observations, there are still different opinions regarding their driving mechanism and physical nature. Large-scale EUV waves were initially believed to be driven by the flare pressure pulses as the chromospheric Moreton waves \cite{1968SoPh....4...30U} (e.g.,\cite{2002A&A...383.1018K,2003SoPh..212..121H,2004A&A...418.1117W}). After the launch of the \em {Solar Dynamics Observatory}~\em (SDO;\cite{2012SoPh..275....3P}) in 2010, the high temporal and high spatial resolution observations taken by the \em Atmospheric Imaging Assembly\em ~(AIA; \cite{2012SoPh..275...17L}) pushed forward the research for EUV waves to a new climax. More and more observational evidence has shown that large-scale EUV waves are closely associated with CMEs rather than flare pressure pulses (\cite{2006ApJ...641L.153C,2012ApJ...753...53S,2012ApJ...754....7S,2013A&A...556A.152X,2012ApJ...753..112Z,2005ApJ...631..604C,2017SoPh..292....7L,2017ApJ...851..101S}). Regarding the physical nature, EUV waves are generally recognized as fast-mode magnetosonic waves driven by  CMEs. Some observations demonstrated the simultaneous existing of two types of waves \citep[e.g.,][]{2004A&A...427..705Z,2011ApJ...732L..20C,2012ApJ...752L..23S,2013ApJ...773L..33S,2014ApJ...786..151S,2012ApJ...753...52L,2015ApJS..219...36G,2018ApJ...863..101C,2013SoPh..288..255K}, as what have been predicted in \cite{2002ApJ...572L..99C,2005ApJ...622.1202C}, i.e., a fast true wave followed by a slow pseudo one. It is worth pointing out that CMEs are not a necessary requirement for producing large-scale EUV waves, although they have been generally accepted as the driver of EUV waves. Some recent observations have shown that large-scale EUV waves can also be driven by various non-CME-association solar eruptions, such as coronal jets \citep{2018MNRAS.480L..63S,2018ApJ...860L...8S,2018ApJ...861..105S,2013ApJ...764...70Z}, mini-filament eruptions \citep{2020ApJ...894...30W}, expanding coronal loops \citep{2015ApJ...804...88S,2017ApJ...851..101S,2018MNRAS.480L..63S,2018ApJ...860L...8S}, and newly formed sigmoidal loops \citep{2012ApJ...753L..29Z}. Recently, large-scale quasi-periodic EUV waves with periods of a few minutes have been detected, which can propagate simultaneously along and across magnetic field lines and cause the oscillations of filaments and coronal loops\citep{2012ApJ...753...52L,2019ApJ...873...22S}. \cite{2019ApJ...873...22S} proposed that the excitation of the large-scale quasi-periodic EUV wave train was possibly due to the period opening of the unwinding helical threads of the associated erupting filament.

A new important finding of the {\em SDO} was the discovery of the so-called quasi-periodic fast propagating (QFP) magnetosonic wave trains \citep{2011ApJ...736L..13L}. QFP wave trains are composed of multiple coherent narrow arc-shaped wavefronts, propagating along open or closed coronal loops with high speeds up to \speed{2000}. Since one or more periods of a QFP wave train are generally consistent with that of the accompanying flares, they are believed to be excited by the periodic energy release process in flares (e.g., \cite{2011ApJ...736L..13L},\cite{2012ApJ...753...53S},\cite{2013SoPh..288..585S},\cite{2013A&A...554A.144Y},\cite{2013A&A...553A.109K},\cite{2017ApJ...844..149K},\cite{2018ApJ...853....1S},\cite{2021ApJ...908L..37M}). Besides, the leakage p-model oscillations from photospheric to the corona \citep[e.g.,][]{2012ApJ...753...53S,2017ApJ...851...41Q}, the dispersive evolution \citep[e.g.,][]{2013A&A...554A.144Y,2014A&A...569A..12N,2018ApJ...853....1S,2018ApJ...860L...8S} and even CMEs \citep{2019ApJ...871L...2M} are also considered as the possible exciter for QFP wave trains. According to \cite{2014SoPh..289.3233L}, QFP wave trains with a series of arc-shaped wavefronts at speeds of \speed{500-2200} propagate upward along funnel- or conic-like coronal loops within a limited angular sector (say, $10^{\circ}\--60^{\circ}$), which indicate that QFP wave trains are trapped in low Alfv\'{e}n structures that serve as waveguides \citep{1983Natur.305..688R,1984ApJ...279..857R}. The lifetimes (periods) of QFP wave trains range from several minutes to more than one hour (25$\--$400 seconds). So far, the direct observations of QFP wave trains are still relatively scarce. However, according to \cite{2016AIPC.1720d0010L}, at least one-third of events associated with large-scale EUV waves, flares, and CMEs are accompanied by QFP wave trains. The lack of a detailed case to analyze QFP wave trains is probable because the QFP waves can only lead to a subtle intensity fluctuation (say, 1\%--5\%) in EUV images. A few two- and three-dimensional MHD simulation works were also performed to study the QFP wave trains \citep{2011ApJ...740L..33O,2013A&A...560A..97P,2016ApJ...823..150T}, and some of them can perfectly reproduce QFP wave trains which have similar amplitude, wavelength, and propagation speeds with real observations \citep[e.g.,][]{2011ApJ...740L..33O}.

In this paper, we report a rare observations of simultaneous detection of three types of fast-mode magnetosonic waves in a prominence eruption event on 2013 April 23. The eruption was accompanied by {\em GOES} C3.0 flare and a broad CME. During the eruption, we observed a large-scale EUV wave ahead of the CME front, a large-scale quasi-periodic EUV wave train inside the CME bubble, and a narrow quasi-periodic EUV wave train trapped by a closed-loop system. The co-existence of the three types of wave phenomena offers an excellent opportunity to understand their driving mechanisms, propagation properties, and the surrounding coronal conditions.

\section{Observations and Data Analysis}
AIA onboard the {\em SDO} provides full-disk (4096 $\times$ 4096 pixels) images of the corona and transition region at seven narrow EUV channels, at a high pixel resolution of 0.$^{\prime\prime}$6, a temporal cadence of 12 seconds, and a wide field of view (FOV) of 1.3 $R_\odot$. The EUV emission lines cover a wide range of temperature from 1 MK to 20 MK, which thus offers a good opportunity for us to study the kinematics and thermal feature of various fast-propagating wave phenomena in the corona. Here we mainly use the images at 171 \AA~(Fe IX; characteristic temperature: 0.6$\times10^6$ K), 193 \AA~(Fe XII,XXIV; characteristic temperature: $1.6\times10^6$ K, $2\times10^7$ K) and 304 \AA~ (He II; characteristic temperature: $0.5\times10^5$ K). All of the AIA images used here are calibrated and differentially rotated to a reference time (18:00:00) with the standard procedure ``aia\_prep.pro'' available in the SolarSoftWare (SSW) package. Since the exposure times of the AIA images vary in response to the increase of flare emission, we normalized it by dividing the exposure time recorded in the fits header of each image. The radio spectrographs provided by the Radio and Plasma WAVE Experiment (WAVES; \cite{1995SSRv...71..231B}) aboard \emph{Wind} spacecraft are used to analyze the associated interplanetary radio signal, which sweeps the frequency range 0.02--1.04 MHz (RAD1) and 1.075--13.825 MHz (RAD2) every minute. We also used the Potential Field Source Surface (PFSS; \cite{2003SoPh..212..165S}) model to show the magnetic loop systems of interest. In addition, Large Angle and Spectrometric COronagraph (LASCO; \cite{1995SoPh..162..357B}) images are used to portray the associated CME caused by the eruption of prominence. Geostationary Operational Environmental Satellite (GOES) 0.5-4 {\AA} flux is used to analyze the periodicity of the accompanying C3.0 flare. To highlight the moving features, for example, the wavefronts, we utilize running-difference and base-difference images to study the kinematics and propagating characteristics of the EUV waves, in which the time difference ($\Delta t$) for running difference images and reference time for base-difference images are 12 seconds and 18:00:00 UT, respectively. To show the kinematic features of the propagating disturbance, we estimated its speed using time-distance stack plots (TDSPs) created by stacking a one-dimensional intensity profile along the center of propagating path of interest, in which the gradient of a oblique stripe represents the propagating disturbance speed. The technique of wavelet analysis (\cite{1998BAMS...79...61T}) is used to investigate the periodicity of the EUV wave trains and flare.

\section{Result}
The event occurred on 2013 April 23 at about 18:00 UT in NOAA active region AR11723 that was close to the west limb of the solar disk from the Earth's angle of view. Such a limb event is in favor of investigating the relationship between the associated CME and EUV waves. An overview of the initial magnetic topology of the eruption source region before the event is shown in Figure 1 (a), in which some extrapolated magnetic field lines are overlaid. We focus on two groups of loop systems: the one is a closed trans-equatorial loop system L1 that connected the active regions of AR11723 and AR11726 in opposite hemispheres, the other one L2 is an active region loop whose two footpoints were rooted in opposite magnetic polarity regions of AR11726. This event started with the rising and eruption of a loop-like prominence, and was associated with a broad CME with an average speed of \speed{403} \footnote[1]{http://cdaw.gsfc.nasa.gov/CME\_list}(see Figure 1 (e)), a {\em GOES} soft X-ray C3.0 flare, a type III radio burst, and various EUV waves. The start, peak, and end times of the flare were at about 18:10 UT, 18:33 UT, and 18:45 UT, respectively. In this paper, we mainly focus on the driving mechanisms of the  accompanying EUV waves, including a large-scale EUV wave traveling ahead of the CME bubble, a quasi-periodic large-scale EUV wave train running upwardly  within the CME bubble (i.e., behind  the CME front), and a narrow quasi-periodic EUV wave train propagating along the closed-loop system L1. In the following subsections, we discuss the driving mechanisms and wave properties of these EUV waves in more detail.

  \begin{figure}    
   \centerline{\includegraphics[width=0.8\textwidth,clip=]{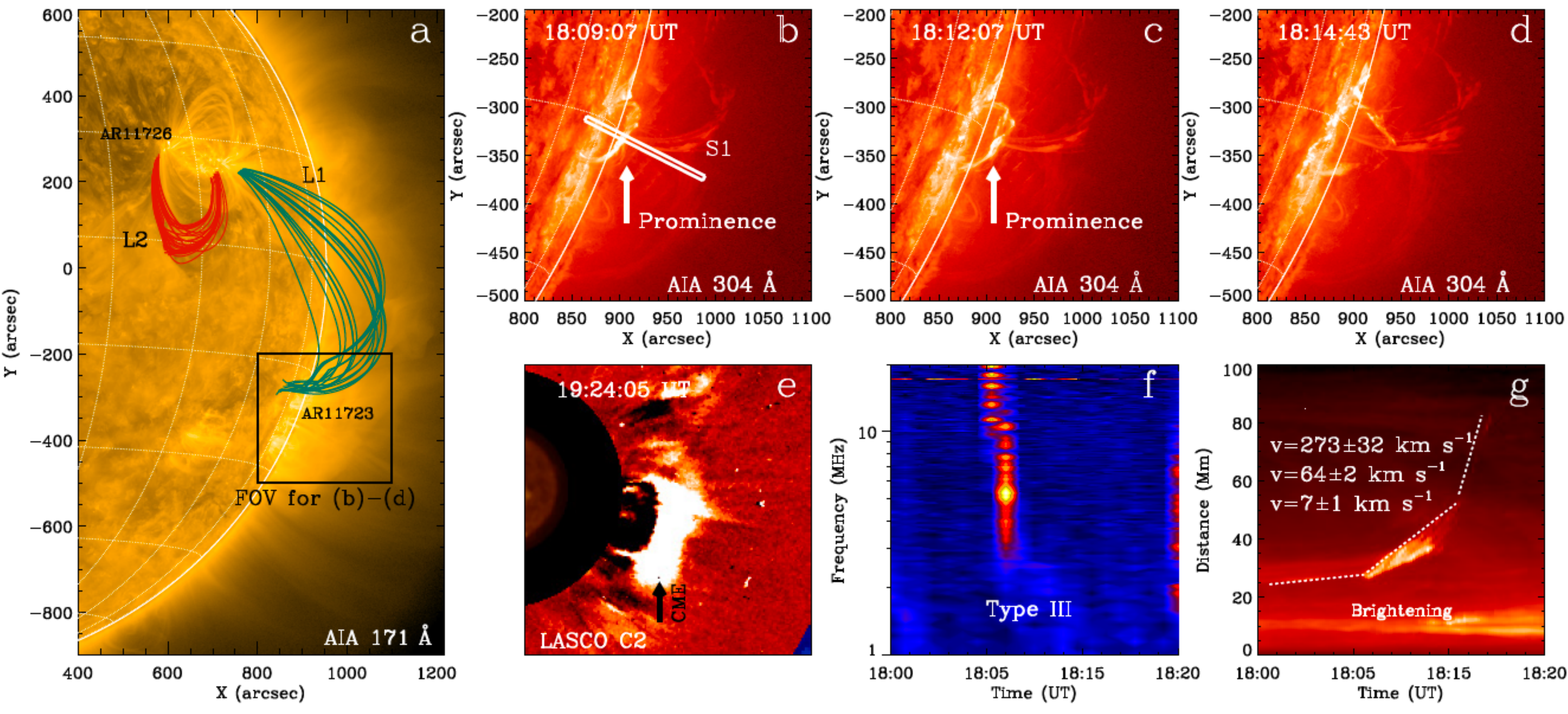}
              }
              \caption{Coronal environment before the eruption in AIA 171 \AA (a) and the evolution of the eruption prominence (b)-(d), as well as the associated CME (e) and  type III radio burst (g).  Time-distance stack plot made from rectangle slice, labeled as ``S1'' in (b), extending from flare epicenter in AIA 304 \AA\ for eruption prominence (g). Note that the white arrows point to the eruption prominence, while the CME front is pointed out by the black arrow. Two groups of loop systems labeled as ``L1'' and ``L2'' in panel (a), are highlighted by extrapolated magnetic field lines.}
   \label{F-simple}
   \end{figure}

\subsection{Prominence Eruption, CME and the CME-Driven EUV wave}

The prominence showed a loop-like shape over the northwest limb; its eruption experienced three distinct phases: slow rising, fast rising, and violent eruption phases (see Figure 1 (b)--(d) and (g), and the animation available in the online journal). The slow rising phase was from 18:00 UT to 18:06 UT, and the prominence rose slowly at a speed of about \speed{7}. The fast rising phase was from 18:06 UT to 18:16 UT, and the prominence rose quickly was about \speed{64}. After 18:16 UT, the prominence erupted violently at an average speed of about \speed{273}. At 18:06 UT, the beginning of the fast rising phase of the prominence eruption, a type III radio burst was detected by the {\em WAVES}, and the prominence body became brighter than before (see Figure 1 (f) and (g)). Generally, type III radio bursts are considered to be a signature of accelerated energetic electrons streaming, channeled along open magnetic field lines, attributed to the reconfiguration of an unstable coronal magnetic field (\cite{2011ApJ...733L..25K,2014RAA....14..773R,2014RAA....14..843G,2015ApJ...807...72L,2017ApJ...851...67S}). The type III radio burst observed here appeared at about 13 MHz, and it rapidly drifted to 3 MHz; this suggests the start of magnetic reconnection in the eruption and the propagation of accelerated energetic electrons from lower corona to higher altitude. In the wake of the ejected prominence, the flare reconnection is believed to occur between two legs of the envelope with a vertical current sheet inside, resulting in the further acceleration upward of prominence (\cite{2001ApJ...552..833M,2006PhRvL..96y5002K,2012ApJ...760...81K}).

     \begin{figure}    
   \centerline{\includegraphics[width=0.8\textwidth,clip=]{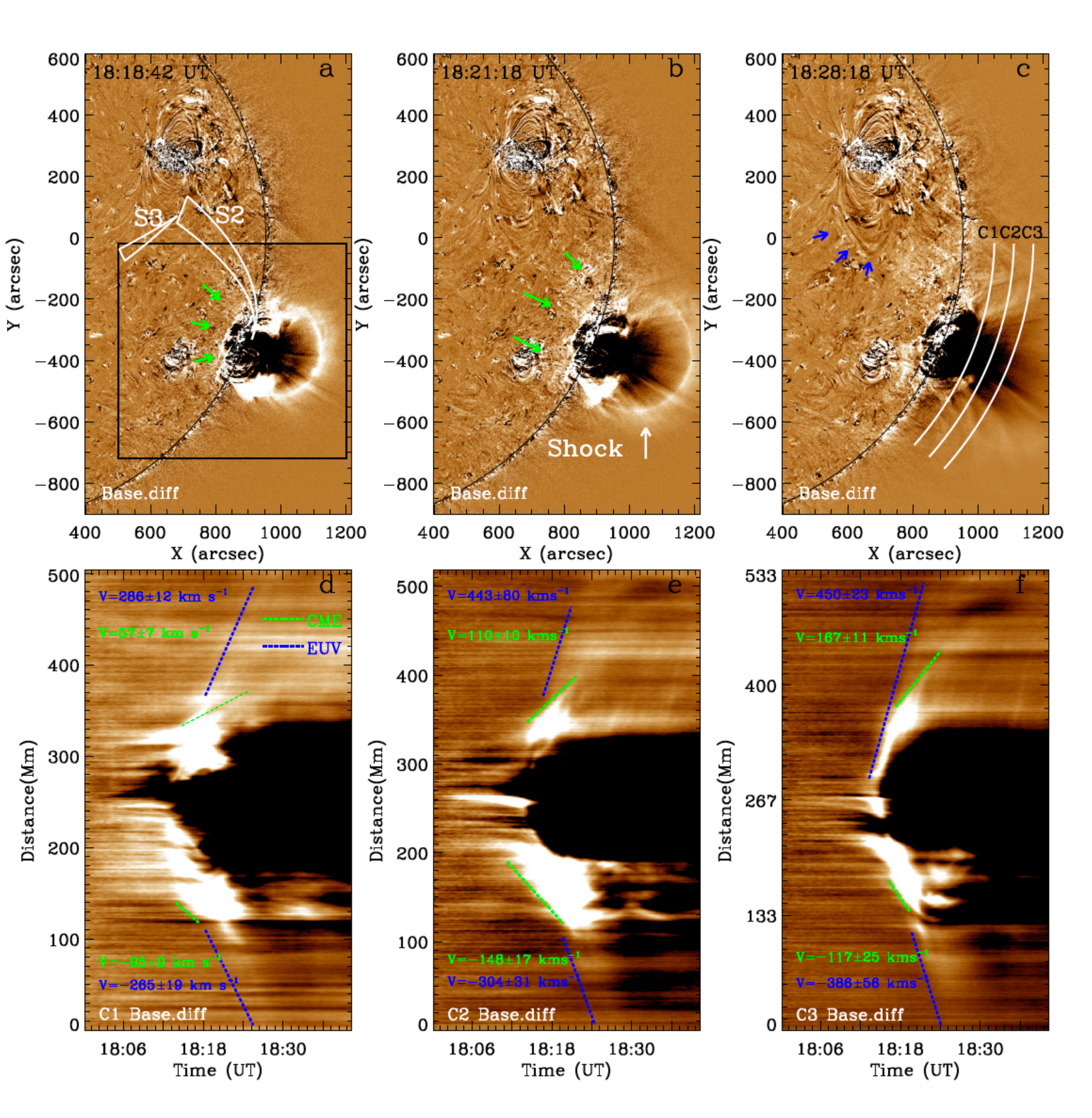}
              }
              \caption{The evolution of the large-scale EUV wave in AIA 193 \AA\ base-difference images (top), and time-distance stack plots (bottom) for the three azimuthal paths labeled as ``C1'', ``C2'', and ``C3'' in panel (c). In panel (b), a shock ahead of the CME bubble is identified by white arrow. In panels (d)-(f), the moving features of the CME bubble and its lateral EUV wave are respectively outlined by the green and blue dashed lines, and their corresponding propagating velocities, determined by applying linear fits, are written with corresponding colors. The green arrows indicate the wavefront of the on-disk propagating EUV wave component, while the blue arrows mark the reflation wavefront. The black box denotes the FOV of  panels (a)-(c) in Figure 4.}
              
   \label{F-simple}
   \end{figure}

During the fast rising phase of the prominence, one can identify the CME bubble in the low corona over the southwest limb of the solar disk as displayed in the top row of Figure 2, which showed as a well-defined bright, sharp circular boundary. A large-scale EUV wave appeared ahead of the CME bubble a few minutes after the formation of the CME bubble (see Figure 2 (b)). In the meantime, diffuse bright wavefront can also be observed on the solar surface (see green arrows in Figure 2 (a) and (b)). The on-disk propagating EUV wave was further interacted with the remote active region AR11726 in the northern hemisphere, after which a reflected EUV wave appeared on the southeast side of AR11726 (see the blue arrows in Figure 2 (c)). Here, both the on-disk and radial propagating EUV waves should be driven by the CME, as those have been reported in previous studies \citep[e.g.,][]{2010ApJ...716L..57V,2012ApJ...752L..23S,2020ChSB}. To analyze the kinematics of the EUV wave ahead of the CME bubble, we selected three azimuthal paths, labeled as  ``C1--C3'' in Figure 2, to generate TDSPs from the AIA 193 \AA\ base-difference images. Here, the azimuthal paths C1--C3 are part of concentric circles centered at the center of the solar disk, and their height above the disk limb are 60, 100, and 140 Mm, respectively. The TDSPs generated along paths C1--C3 are displayed in Figure 2 panels (d)--(f). In each of TDSPs, the boundary of the CME bubble can be identified as a paired bright bidirectional oblique stripes, and they are indicated by the green dotted lines. Based on these TDSPs, the lateral expansion speeds in south and north directions are measured to be \speed{95--148} and \speed{57--167}, respectively. The EUV wave ahead of the CME bubble can well be identified in these TDSPs, they are observed fainter than the CME bubble. The EUV wave firstly mixed with the CME bubble, it became clear until the expansion of the CME bubble started to slow down at about 18:18 UT. The generation of the EUV wave was directly separated from the outer edge of the CME bubble, and it propagated at a speed of about \speed{265--450}, which is faster than the CME bubble about 2--4 times in the low corona. These observational results suggest that the observed EUV wave ahead of the CME bubble was driven by the expansion motion of the CME bubble, consistent with previous observations \citep[e.g.,][]{2011ApJ...738..160M,2012ApJ...753...52L,2012ApJ...745L...5C,2019ApJ...871....8L}.

  \begin{figure}    
   \centerline{\includegraphics[width=0.8\textwidth,clip=]{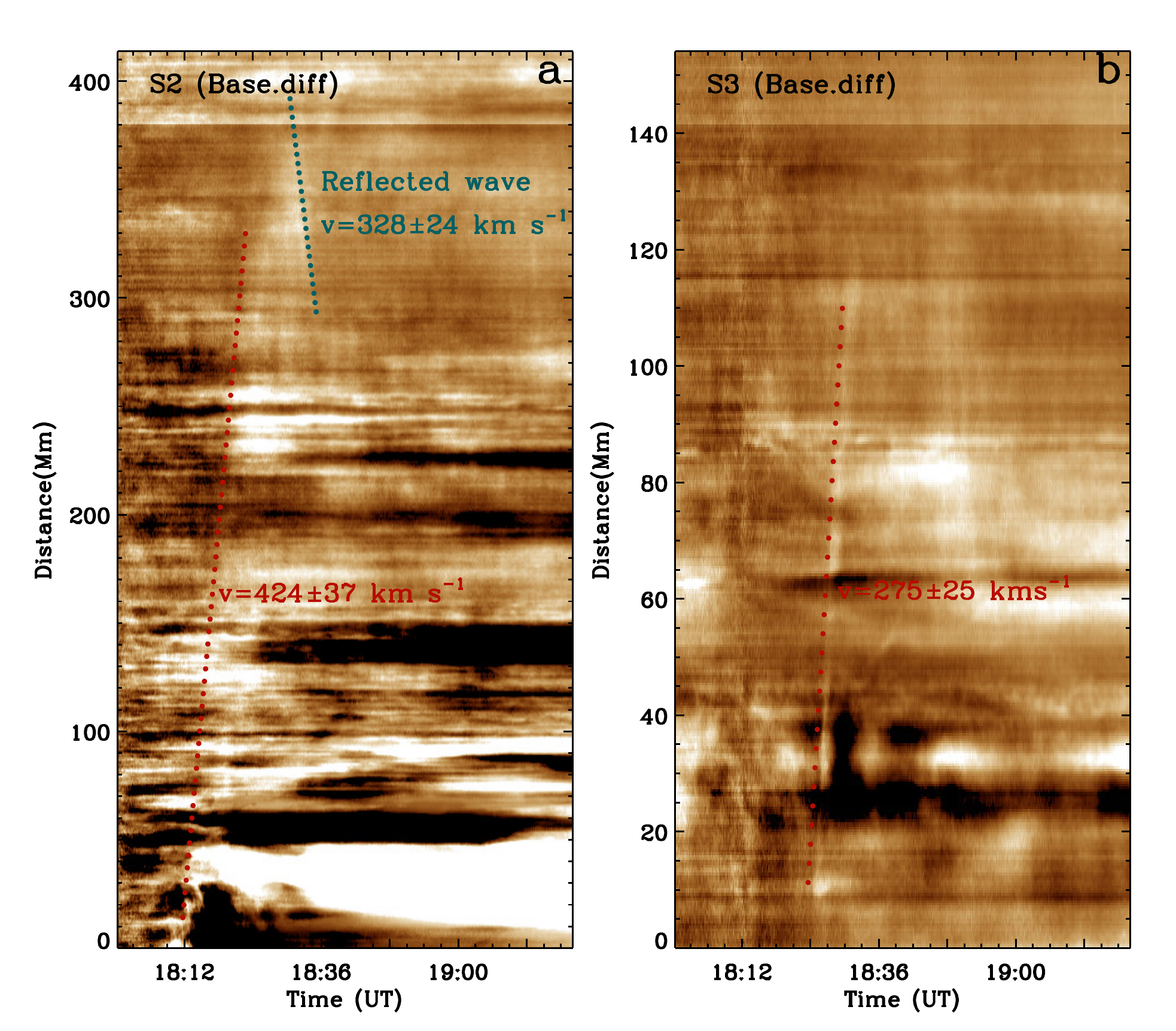}
              }
      \caption{ 
      Time-distance stack plots for the two sectors, labeled as ``S2'' and ``S3'' in Figure 2, in AIA 193 \AA\ base-difference images. The velocities of the wavefronts are determined by using linear fits and they corresponding speeds are written around them. 
      }
   \label{F-simple}
   \end{figure}

The on-disk propagating EUV wave component showed a semicircular shape and propagated oriented to the northeast direction. It interacted with AR11726 and produced a reflected wave propagated oriented to the southeast direction. To quantify the kinematics of these on-disk propagating EUV waves, we generated two TDSPs along sectors S2 and S3 as indicated in Figure 2 (a). We use AIA 193 \AA\ base-difference images to generate the TDSPs, and the results are displayed in Figure 3. Note that the origin of sector S2 is located near the flare epicenter and orient to the loop system L2, while S3 is along the propagation direction of the reflected wave. In the TDSP along S2 (Figure 3 (a)), one can identify the start time of the on-disk propagating EUV wave was about 18:12 UT. The EUV wave showed obvious deceleration during the propagation, and the average projection speed is measured to be about \speed{424}. At 18:24 UT, the EUV wave interacted with AR11726 and produced a reflected wave that propagated in the opposite direction of the primary EUV wave. Along S2, the reflected EUV wave had a speed of about \speed{330}, significantly slower than the incident EUV wave. To more precisely measuring the speed of the reflected wave, we generate a new TDSP along S3 (Figure 3 (b)). This path is perpendicular to the propagation direction of the reflected wave. It can be seen that the reflected wave also showed a deceleration, and had an average speed of about \speed{275}.

\subsection{Flare-Ignited Quasi-Periodic EUV Wave Trains}

Besides the single pulse EUV waves ahead of the CME bubble and on the solar surface, it is striking that multiple successive arc-shaped wavefronts were observed within the CME bubble during the impulsive phase of the flare. We show this large-scale quasi-periodic wave train in Figure 4 with AIA 193 \AA\ running difference images (see the white arrows and the animation available in the online journal). The wavefronts emanated continuously from the flare epicenter and exhibited as a series of concentric semicircles resembling the shape of the CME bubble; they propagated outwardly and gradually disappeared when they reached up to the height of the boundary of the CME bubble. Using the AIA 193 \AA\ running difference images, we made a TDSP along the rectangle in Figure 4 (a) to study the kinematics of the large-scale quasi-periodic wave train inside the CME bubble, and the result is displayed in Figure 4 (d). In this TDSP, the brightest stripe is the boundary of the CME bubble, and the EUV wave ahead of the CME can be identified as a relatively fainter stripe on the left of the CME one. The large-scale quasi-periodic EUV wave train can be recognized as a series of weak stripes on the right of the CME one. By applying quadratic fit to these stripes, we obtain the average speed of the CME bubble and the large-scale quasi-periodic wave train were about \speed{300} and \speed{1100}, respectively. The deceleration and period of the large-scale quasi-periodic wave are estimated to be $\alpha=$ \acc{-4.1} and 120 seconds, respectively. Here, we note that the speed of the large-scale quasi-periodic wave train was significantly faster than the CME bubble and the single EUV wave ahead of the CME. In addition, the large-scale quasi-periodic wave train was behind of the CME bubble, and its appearance time was delayed about 4 minutes with respect to that of the rapidly rising phase of the CME bubble. Therefore, the large-scale quasi-periodic wave train was unlikely driven by the CME expansion as those have been documented in the literature \citep[][]{2015LRSP...12....3W,2020ChSB}.

\begin{figure}
\centerline{\includegraphics[width=0.8\textwidth,clip=]{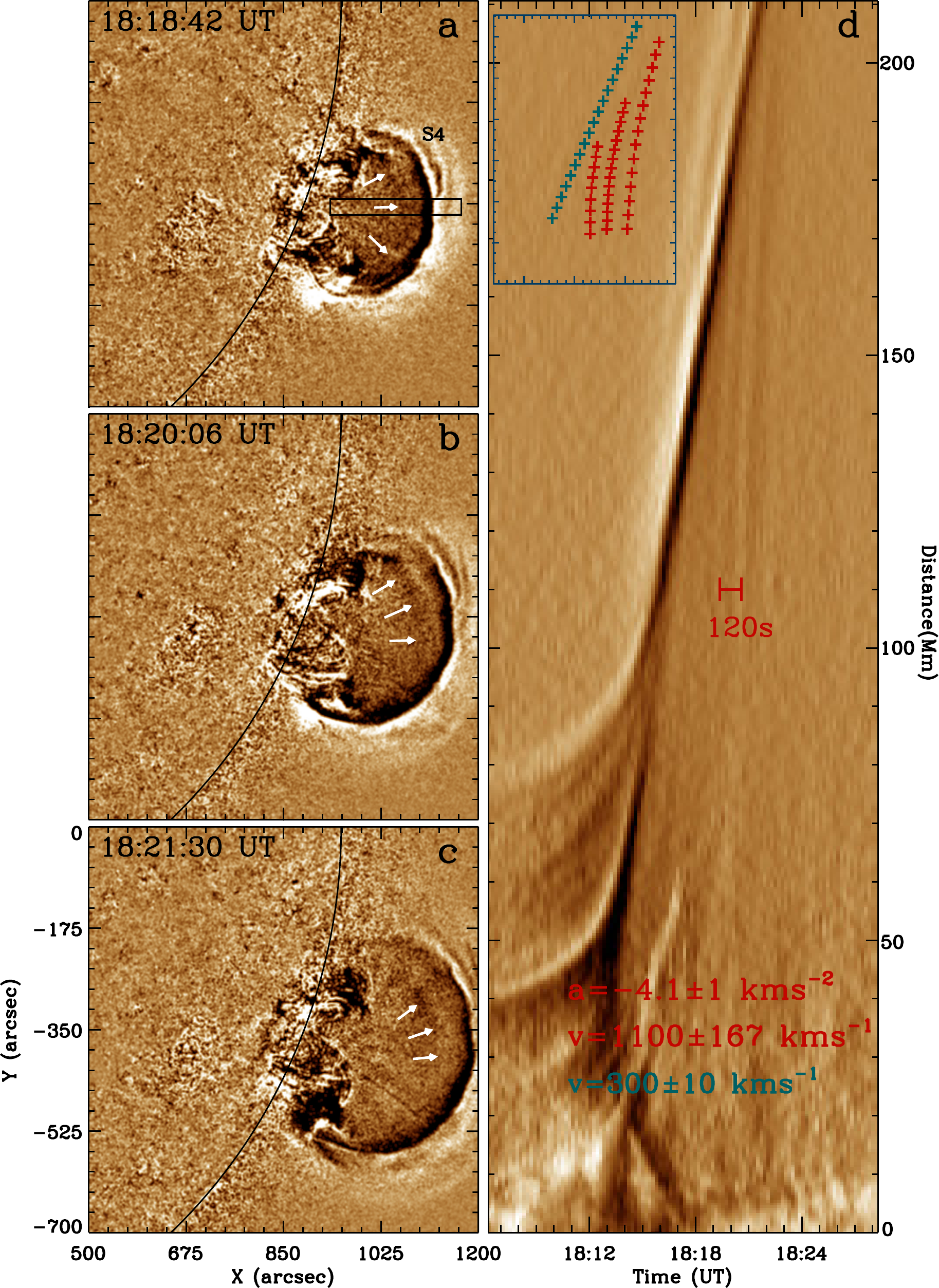}}
\caption{Evolution of the large-scale quasi-periodic EUV wave running inside the CME bubble. Panels (a)-(c) show the evolution of wavefront in AIA 193 \AA\ running-difference images, in which the white arrows point to the wavefronts. The FOV of panels (a)-(c) are the region indicated by the black box in Figure 2 (a).  Panel (c) shows the AIA 193 \AA\ running-difference time-distance stack plot along a rectangle slice, labeled as ``S4'' in panel (a). The kinematic parameters of the CME bubble and the wavefronts of the large-scale quasi-periodic EUV wave are get from quadratic fits of the brighten ridges, labeled in panel (d). To show the wavefronts clearly, we outline their tracks  into the top-left corner of panel (d). 
}
\label{F-simple}
\end{figure}

Along the closed-loop system L1, we observed another quasi-periodic fast propagating EUV wave train that can well be identified in the AIA 171 \AA\ running difference images (see Figure 5 (a)--(c) and the animation available in the online journal). Compared with the large-scale quasi-periodic EUV wave train inside the CME bubble, this narrow quasi-periodic EUV wave train also emanated continuously from the flare epicenter, but it had a relatively smaller angular extent and its propagation was obviously guided by the coronal loop rather than confined within the CME bubble. These observational characteristics make us believe that the two observed quasi-periodic EUV wave trains were two independent EUV wave trains, but both of their generations were possibly associated with the flare. Here, the narrow quasi-periodic EUV wave train should be a so-called QFP magnetosonic wave as those reported in previous studies \citep[e.g.,][]{2011ApJ...736L..13L,2013A&A...554A.144Y,2012ApJ...753...53S,2013SoPh..288..585S,2021ApJ...908L..37M}. The propagations of such QFP waves are typically trapped in coronal loops acting as waveguides (high density, low Alfv\'{e}n speed). Generally, QFP magnetosonic waves often appear during the impulsive phases of the associated flares. In the present case, the narrow quasi-periodic EUV wave train can be observed from about 18:24 UT to 19:12 UT, lasting up to 50 minutes from the flare impulsive phase to the decay phase. The TDSPs are generated by using the AIA 171 \AA\ and 193 \AA\ images along the S5 as indicated in Figure 5 (a), and the result are displayed in Figure 5 (d) and (e), respectively. In AIA 171 \AA\ TDSP, one can identify many bright parallel stripes that represent the propagating wavefronts of the QFP wave train. These wavefronts can also been found in AIA 193 \AA\ TDSP, but they are relatively fainter than it detected in AIA 171 \AA\, (see Figure 5 (d)(e) and the animation in the accompanying online material). By applying a linear fit to the stripes, we obtain the average speeds of the narrow quasi-periodic EUV wave train was about \speed{475}.  

  \begin{figure}   \centerline{\includegraphics[width=0.8\textwidth,clip=]{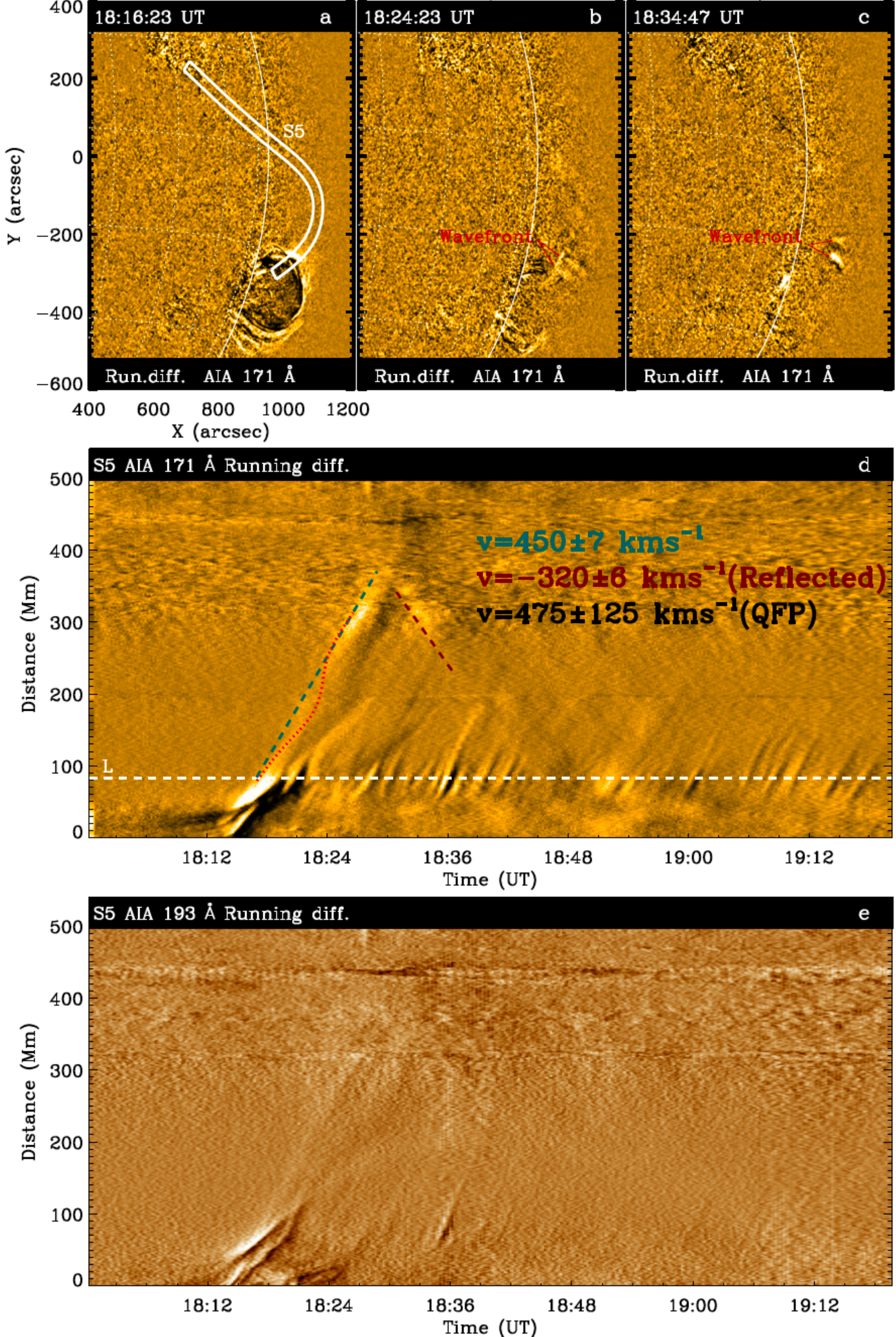}
              }
              \caption{ The evolution of the narrow quasi-periodic EUV waves running along L1 in AIA 171 \AA\ running-difference images (a)-(c). The red arrows point to the wavefronts. Panels (d) and (e) respectively show the time-distance stack plots constructed from AIA 171 \AA\ and 193 \AA\ running-difference images along the white slice labeled as ``S5'' in panel (a), in which the horizontal white dashed line indicates the position whose intensity profile is used to analyze the periodicity of the narrow quasi-periodic wave trains, while the red and blue dashed lines denote the linear fits to the wave ridges.  The red dotted curve line roughly indicates the trend of ridge.  The average speeds of the incident and reflected EUV wave are 450$\pm$\speed{7} and 320$\pm$\speed{6}, respectively. The average velocity of the narrow quasi-periodic EUV waves is 475$\pm$ \speed{125}  }
   \label{F-simple}
   \end{figure}

The first strong stripe in Figure 5 (d) represents the EUV wave ahead of the CME bubble, which was trapped by the closed-loop system L1. It propagated along the entire loop system and produced a reflected wave during its interaction with the remote footpoint of the loop system. The EUV wave propagated at an average speed of about \speed{450}; it was reflected at the remote footpoint of the loop at about 18:27 UT. The reflected wave propagated at a speed of about \speed{320}. It is noted that the generation of the reflected wave along the loop system delayed the on-disk reflected EUV wave by about 3 minutes, although its average speed was faster than that of the on-disk EUV wave. A alternative explanation could be that the EUV wave trapped in the loop propagated a longer distance, as the loop system should be a curving structure. Generally, EUV waves show broadening wavefronts and decreasing amplitudes during their propagation. For the present case, the intensity amplitudes and propagation speeds of the EUV wave at different sections of the loop showed an abnormal varying pattern, i.e., the amplitude and speed were firstly decreasing continuously from the origin to the middle section of the loop, then they followed an increasing trend from the middle section to the remote end of the loop system (see Figure 5 (d)). These features are typical characteristics of large-scale fast-mode magnetosonic waves steepened to form freely propagating shocks traveling along loops \citep{1995JPlPh..53..109M}. The intensity amplitude of an EUV wave is determined by the energy and geometric of the waveguide, and the propagating speed is related to the amplitude \citep{2010AdSpR..45..527W,2015LRSP...12....3W}. Obviously, the middle section of the loop system had a broader cross-section than that near the two footpoints. As the EUV wave propagated from one end of the loop to the middle section, the wave energy spread to a broader extent and therefore led to a decrease of the intensity amplitude \citep{2013A&A...554A.144Y}. During this period, the EUV wave evolved from a shock or non-linear fast-mode magnetosonic wave into a linear fast-mode magnetosonic wave. From the middle section of the loop to the other end, the wave energy and amplitude experience a reverse process, and the linear fast-mode magnetosonic wave became a non-linear fast-mode magnetosonic wave or shock close to the remote end of the loop. The speed of a fast-mode magnetosonic wave is directly proportional to the magnetic field strength of the medium in which it propagates. For the closed-loop system L1, the magnetic field strengths close to the two ends should be greater than that close to the apex. Therefore, the wave speed was naturally experienced first decreasing and then increasing processes. In practice, the speed of the fast-mode magnetosonic wave is also determined by plasma density and sound speed, as well as the projection effect can also be important. Therefore, the actual measured speed of the EUV wave was inflected for many reasons, such as the inhomogeneous distribution of the plasma density and magnetic field strength, local sound speed, and projection effect.

\subsection{Periodicity of the Quasi-Periodic EUV Wave Trains and the Flare}

\begin{figure}
\centerline{\includegraphics[width=0.8\textwidth,clip=]{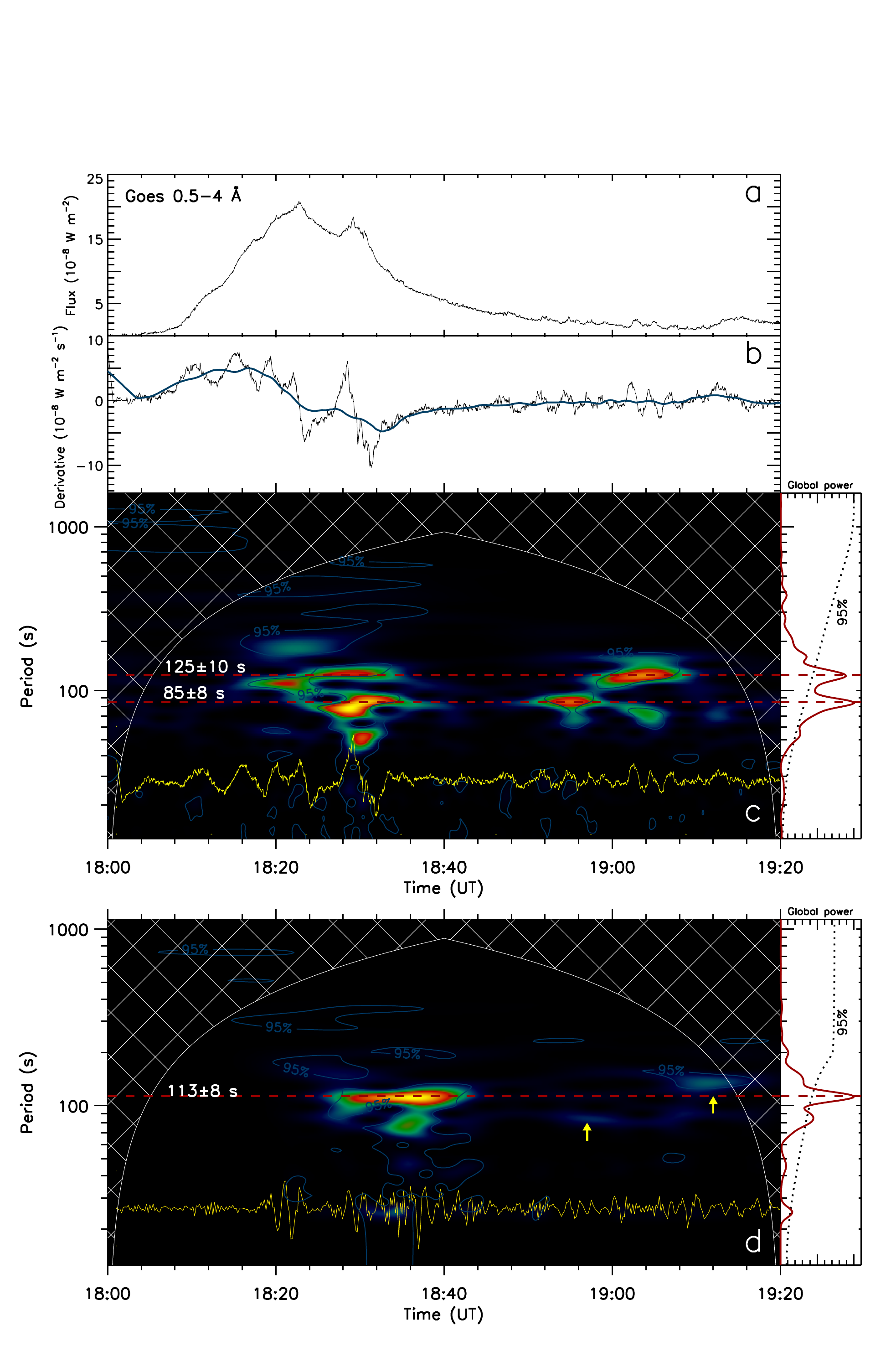}}
\caption{Periodicity analysis for the narrow quasi-periodic EUV  wave and the associated flare pulsation. Panels (a)-(b) are {\em GOES} 0.5-4 \AA\ soft X-ray flux and its time derivative, respectively. The yellow curve in panel (c) is smooth/detrended time derivative signal of the {\em GOES} 0.5-4 \AA\ soft X-ray flux after subtracting the blue smooth curve, while the yellow curve in panel (d) is the smooth/detrended intensity profile of the horizontal white dashed line labeled as ``L'' in Figure 5 (d), and their corresponding wavelet power spectrums are displayed in panel (c) and (d), respectively. The dotted line in global power indicates the 95\% significance level, while the red horizontal dashed lines mark the periods.}
\label{F-simple}
\end{figure}

The periodicity of the quasi-periodic EUV wave trains and the associated flare is analyzed in detail by using the wavelet analysis method \citep{1998BAMS...79...61T}. Figure 6 (a) displays the {\em GOES} soft X-ray flux profile in the energy band of 0.5--4 \AA. Since this flare lacks hard X-ray observations and limb occultation of the flare ribbons in the UV and EUV observations, we use the time derivative of the {\em GOES} 0.5--4 \AA\ soft X-ray flux to analyze the periodicity of the flare. According to the Neupert effect \citep{1968ApJ...153L..59N}, such time derivative soft X-ray fluxes can be viewed as a proxy of the corresponding hard X-ray fluxes that reflect the variation of non-thermal particles accelerated by the magnetic reconnection in flares. Figure 6 (b) shows the time derivative curve of the {\em GOES} 0.5--4\AA\ soft X-ray flux during the flare. We subtracted a 400 seconds smoothed curve (blue) to detrend the signal and the result is overlaid in Figure 6 (c). Using this signal as input, we generated the wavelet spectrum, and the result is displayed in Figure 6 (c). One can identify several main periods in the wavelet spectrum, they appeared mainly during two time spans of 18:10--18:40 UT and 18:50--19:15 UT. It is determined based on the global power of the wavelet spectrum map that the main dominated periods in the flare were about 85 and 125 seconds, which is similar to the result analyzed the hard X-ray \citep{2017A&A...597L...4L}.

Since the wave signal of the large-scale quasi-periodic EUV wave train inside the CME bubble was too weak, we failed to obtain its period with the wavelet technique. As indicated by the short red bar in Figure 4 (c), we directly estimated its period to be about 120 seconds by measuring the time span during two neighboring wave crests. The wave signal of the narrow quasi-periodic EUV wave train was relatively stronger. Therefore, we analyzed its periodicity by using the intensity profile varying with time, which can be obtained by extract the intensity profiles along the horizontal lines marked ``L'' as shown in Figure 5 (d). We also use the detrended intensity profile as the input for the wavelet procedure; it is also overlaid in the wavelet spectrum map as a yellow curve in Figure 6 (d). In the same way, we determined the periods of the EUV wave train based on the peak of the global power curve of the wavelet spectrum. It is obtained that the main period of the narrow quasi-periodic EUV wave was about 113 seconds, which is roughly similar to the result reported by \cite{2017ApJ...844..149K}. Obviously, all the obtained periods of the large-scale and narrow quasi-periodic EUV wave trains were the same with those of the pulsations in the flare, suggesting their common origin.

\subsection{Wave energy and seismology application}
The energy flux $\mathcal{F} $ per unit area and time carried by the waves can be estimated from the kinetic energy of the perturbed plasma using the expression $\mathcal{F} =\frac{1}{2}\rho (\delta v)^2v_{ph}$, where $\rho$ is the mass density, $v$ is the disturbance speed of local plasma, and $v_{ph}$ is the phase speed of the EUV wave \citep{2004psci.book.....A}. It is generally believed that the observed EUV emission intensity is proportional to the square of the plasma density, i.e., $ I \varpropto \rho^2$. Thus the density modulation $\delta \rho$ and the observed emission intensity variation $\delta I ~(\delta I=(I-I_0)/I_0)$ have the relation $\delta \rho/\rho = \delta I /(2I)$, where $I$ and $I_0$ respectively are highest amplitude and background intensity. Following \cite{2011ApJ...736L..13L} and assuming that the observed intensity variation $\delta I$ stems from the density modulation $\delta \rho$ and using $\delta v /v_{ph} \geqslant \delta I/(2I)$, then the energy flux $\mathcal{F} $ of the perturbed plasma could be rewritten as 
$\mathcal{F} \geqslant\frac{1}{8}\rho(\delta I/I)^2 v_{ph}^3$. Here we assume that the electron-number density of the corona and the waveguiding loops are $n_e= 1.5\times10^8 $ cm$^{-3}$ \citep{2004psci.book.....A} and $n_e= 6\times10^8 $ cm$^{-3}$ \citep{2004ESASP.575...97A}, respectively. Meanwhile, we take the average phase speeds for the observed large-scale EUV wave, the large-scale quasi-periodic EUV wave train, and the narrow quasi-periodic EUV wave train  are \speed{424}, \speed{1100} and \speed{475}, and their corresponding relative intensity perturbations of $\delta I$ are 14\%, 15\% and 5\%, respectively. Then we arrive at the energy fluxes  $\mathcal{F}$ for the three types of EUV waves are $0.6\times 10^5$ erg cm$^{-2}$ s$^{-1}$, $12.0\times 10^5$ erg cm$^{-2}$ s$^{-1}$ and $0.43\times 10^5$ erg cm$^{-2}$ s$^{-1}$, respectively. With the same method, we calculate the energy flux of the other waves. The corresponding parameters used in our calculation and the obtained results are all listed in table 1. We find that the energy flux density sorted in decrease turn as followed: Large-scale quasi-periodic EUV wave train, EUV wave along the closed-loop, reflected wave along the closed-loop, EUV wave on the solar surface, narrow quasi-periodic EUV wave train, on-disk reflected wave. At the meantime, the energy density of reflected waves are significant smaller than their corresponding incident waves. This is probably due to the loss in energy during the interaction, or only a part of the incident waves were reflected. It is noted that the energy flux density of the trapped EUV wave along the closed-loop is about seven times of that on the solar surface. Because the two EUV waves were all excited by the expansion of the CME bubble, they should have a similar energy flux density during their initial stage. The large difference between the two EUV waves may suggest that the energy loss rate of the EUV wave along the closed-loop was slower that that on the solar surface, i.e., the loop waveguide can effectively maintain the wave energy during the propagation. In addition, the energy flux density of the large-scale quasi-periodic EUV wave train inside the CME bubble is thirty times of the narrow quasi-periodic EUV wave along the closed-loop system, indicating their distinctively different properties of the two types of EUV wave trains.

\begin{table*}
	\centering
	\caption{Physical Parameters of the Various EUV Waves}
	\label{tab:parameter}
\resizebox{\textwidth}{14mm}{

	\begin{tabular}{lccccc}
	\hline

Various EUV Waves & $v$[ km s$^{-1}]$ & $I/I_0$ & $\delta I$ & $M$ & $\mathcal{F} $ [erg cm$^{-2}$ s$^{-1}]$\\
 
\hline 
EUV wave on the solar surface & 424 & 1.14 &  14\%  & 1.05 & $0.6\times10^5$ \\ 
On-disk reflected wave  & 275 & 1.08 &  8\%  & 1.03 & $0.05\times10^5$ \\
EUV wave along the closed-loop & 450 & 1.17 & 17\% &  1.04 & $ 4.18 \times10^5$\\
Reflected wave along the closed-loop &  320 & 1.15 & 15\% &  1.03 & $ 1.2 \times 10^5$\\ 
\hline
Large-scale quasi-periodic EUV wave train & 1100 & 1.15 & 15\% & 1.05 &$ 12.0\times10^5 $\\
Narrow quasi-periodic EUV wave train  & 475 & 1.05 & 5\%& 1.01 & $0.43\times10^5$ \\
\hline
\end{tabular}}
\end{table*}

\begin{table*}
	\centering
	\caption{Event Timeline }
	\label{tab:time}
	\resizebox{\textwidth}{14mm}{
	\begin{tabular}{ll}
	\hline

  18:00 UT   & Slow rising of the prominence \\ 
  18:06 UT   & Fast rising of the prominence \\ 
  18:12 UT   & Start of the EUV wave ahead of the CME bubble \\
  18:16 UT   & Appearance of the large-scale quasi-periodic EUV wave train inside the CME bubble\\
  18:20 UT   & Appearance of the narrow quasi-periodic EUV wave train along the closed-loop \\
  18:24 UT   & Impingement of the on-disk EUV wave upon the remote active \\
                    & Appearance of the reflected EUV waves on the surface and along the closed-loop \\

\hline
  
	\end{tabular}}
\end{table*}

In the low $\beta$ (the ratio of plasma pressure to magnetic pressure) corona, one can estimate the Mach number of fast-mode magnetosonic waves based on the measurable emission density jump parameter by assuming $I \varpropto \rho^2$, i.e., $X= \frac{\rho}{\rho_0}=\sqrt{\frac{I}{I_0}}$ \citep{1982soma.book.....P}. The expression for the Mach number calculation for a fast-mode magnetosonic wave propagating perpendicular (along) coronal loops can be written as $M = \sqrt{\frac{X(X+5)}{(2(4-X))}}$ ($M = X^{\frac{1}{2}}$). Based on the measured emission density jump values of the EUV waves observed in the present event, we obtained their Mach numbers (see Table 1). It can be seen that the Mach numbers of various EUV waves are in the range of 1.01--1.05, indicating that these waves are all weakly shocked fast-mode non-linear magnetosonic waves. The Mach number for the EUV wave along the solar surface is about 1.05, which for the EUV wave trapped in the closed-loop is about 1.04. It is noted that the Mach numbers of the reflected EUV waves on the solar surface and along the closed-loop are relatively smaller than their corresponding incident waves. For the quasi-periodic EUV wave trains, the Mach number of the large-scale quasi-periodic EUV wave train inside the CME bubble is significantly larger than that of the narrow quasi-periodic EUV wave train along the closed-loop.

With the observational phase speed of the propagating waves described above, we can estimate the local magnetic field strengths of the close-loop and quiet-Sun corona where the EUV waves propagated. The sound speed $c_s$ can be calculated from the formula $c_s=147\sqrt{\frac{\rm T_e}{\rm 1MK}}$ \citep{2004ESASP.575...97A} in a fully ionized plasma. For the formation temperature ($T_e$) of the AIA 193 \AA\ channel, which is at 1.6 MK. Therefore, we can obtain the corresponding sound speed ($c_s$) is about \speed{185}. Based on the formula $v=\sqrt{{c_s}^2+{v_A}^2}$ and the measured speed of the EUV wave (\speed{424}), we can estimate that the Alfv\'{e}n speed and the magnetic strength in the quiet-Sun corona are respectively about \speed{382} and 2.3 Gauss, by assuming that the wave-vector of the on-disk propagating EUV wave was perpendicular to the magnetic field. Here, $v_A$ is the local Alfv\'{e}n speed, which is written as $v_A=\frac{B}{\sqrt{4\pi\rho}}=\frac{B}{\sqrt{4\pi\mu m_p 1.92 n_e}} \footnote[2]{$\mu=1.27$ is the mean molecular weight in the corona, $m_p = 1.67\times 10^{-24}$g is the proton mass.}$\citep{1982soma.book.....P}.

We can estimate the magnetic field strength of the closed-loop system L1 based on the trapped EUV wave and the narrow quasi-periodic EUV wave train, respectively. For a fast-mode magnetosonic wave propagating along the magnetic field, its speed should be the same with the local Alfv\'{e}n speed. So, we can obtain the magnetic field strength of the coronal loop based on the formula $v_A=\frac{B}{\sqrt{4\pi\rho}}$. However, as that had been pointed out in \cite{2019ApJ...873...22S}, the result directly obtained by using the measured speed of EUV wave should be larger than the true value, since the observed EUV waves should be shocks or non-linear fast-mode magnetosonic waves. To derive the real magnetic field of the coronal loop, one should first obtain the linear fast-mode magnetic speeds by dividing the measured speeds of the EUV waves by their Mach numbers. The Mach number of an EUV wave can be calculated by utilizing the formula $M=\sqrt{\frac{n}{n_0}}$  \citep{1982soma.book.....P}, where $n$ and $n_0$ are the peak and background densities. We firstly measured the highest emission intensity amplitude ($\frac{I}{I_0}$) of the EUV wave (1.17) and the narrow quasi-periodic EUV wave train (1.05) along the closed-loop system L1, then we can obtain the Mach numbers of the two EUV waves are respectively about 1.04 and 1.01 by assuming $I \varpropto \rho^2$ and the wave-vectors of the EUV waves were parallel to the magnetic field. The average speed of the EUV wave (narrow quasi-periodic wave train) was \speed{450 (475)}, based on which we finally obtain the magnetic field strength of the closed-loop system is about 5.2 (5.6) Gauss, consistent with the result estimated by \cite{2019ApJ...873...22S}.

\section{Conclusion and Discussion} 
Using high spatiotemporal resolution observations taken by the {\em SDO}, we studied three types of EUV waves launched in a single solar eruption accompanied by a {\em GOES} C3.0 flare, a prominence eruption and a CME. The waves include large-scale single pulse EUV waves propagated simultaneously ahead of the CME bubble and on the solar surface, a large-scale quasi-periodic EUV wave train inside the CME bubble, and a narrow quasi-periodic EUV wave train along a transequatorial closed-loop system. Based on our analysis results, we propose that all these waves are fast-mode magnetosonic waves, in which the large-scale single pulse EUV waves propagated ahead of the CME bubble and on the solar surface were excited by the expansion of the CME, while the large-scale quasi-periodic EUV wave train inside the CME bubble and the narrow quasi-periodic EUV wave train along the closed-loop were probably launched by the intermittent energy releasing process in the flare. To make clear the time relationship among various eruption features of the event, we list the start times of each eruption step in Table 2.

Large-scale single pulse EUV waves associated with on-disk solar eruptions are generally mixed with the associated CME bubbles, thus it is difficult to distinguish their origin. Therefore, direct observational evidence for determining the relationship between EUV waves and CMEs is limited \citep{2017ApJ...851..101S}. Alternatively, limb events are more suitable for such kind of study \citep{2011ApJ...738..160M,2012ApJ...745L...5C}. The present limb event provides a rare opportunity for us to diagnose the relationship between the EUV wave and the CME. It is demonstrated that the EUV wave ahead of the CME bubble was directly separated from the CME bubble during its fast expansion stage, and the formed EUV wave was faster than the CME 2--4 times at different heights. The observation also indicates that the formation height of the EUV wave was about 100 Mm above the solar surface, consistent with previous stereoscopic observations \citep{2014SoPh..289.2565D}. Taken these observational results together, we propose that the rapid CME expansion compresses the surrounding plasma, resulting in the formation of the observed EUV wave \citep{2012ApJ...753...52L,2012ApJ...745L...5C}. Note that we did not detect the slower ``wave'' trailing behind of the fast EUV wave as predicted in \cite{2002ApJ...572L..99C}. Based on the speed relationship between the EUV wave and the associated CME, we tend to think that the CME bubble may correspond to the so-called slower pseudo wave component. EUV waves often interact with remote strong magnetic structures such as active regions or coronal holes, and wave effects such as reflection, refraction, and transmission can be expected at their boundaries \citep{2009ApJ...691L.123G,2012ApJ...756..143O,2013SoPh..288..255K,2015ApJ...803L..23K,2012ApJ...754....7S,2012ApJ...752L..23S,2013ApJ...773L..33S,2019ApJ...873...22S}. The present event also showed a reflected EUV wave when the on-disk propagating EUV wave interacted with the active region AR11726. Its lifetime and traveling distance before disappear were about 6 minutes and 120 Mm, respectively. Besides, the large-scale EUV wave ahead of the CME bubble was partially trapped by the closed-loop system L1, it also showed reflected effect at the remote footpoint. The speeds of the reflected waves in the loop and on the solar surface were all decreased significantly with respect to their corresponding incident waves; this is possibly due to the loss of energy during the interaction. For the trapped EUV wave, it showed a changing intensity amplitude and speeds during its propagation along the loop, which reflects the variation of the physical parameters of the loop system. According to the conservation law of wave energy, we propose that the deceleration (acceleration) of the EUV wave is the result of energy disperse (concentrate) into a broad (narrow) cross-section waveguide, which leads to the increase (decrease) of the intensity amplitude.

The large-scale quasi-periodic EUV wave train observed inside the CME bubble had a projection speed of about \speed{1100}, a period of about 120 seconds, and an intensity amplitude of about 15\% higher than the background. The arc-shaped wavefronts of the wave train were observed to be generated continuously from the flare epicenter and propagated upwardly in the radial direction; they became prominent in a higher altitude at a height greater than 50 Mm from the solar surface, indicating their increasing amplitude since their origin. This seems to imply that we just observed the initial steeping stage of a typical fast-mode magnetosonic wave or shock \citep{2007LNP...725..107W}. Since the wavefronts were inside the CME bubble and propagated radially as the CME bubble, they continuously disappeared when they reached up to the height of the frontal of the CME. Since the speed, amplitude, and intensity profile of the wave train are similar to those of typical large-scale EUV waves, we suggest that it should be a fast-mode magnetosonic wave.

How did the large-scale quasi-periodic EUV wave train originate? In the case of the piston-driven shock, the wavefront should be initially located ahead of the driver and then decouples from the driver and propagates freely in the supporting medium. In most studied cases, CMEs are proposed as the typical driver of large-scale single pulse EUV waves \citep{2014SoPh..289.3233L,2015LRSP...12....3W,2020ChSB}, such as the one ahead of the CME bubble in the present event. For the large-scale quasi-periodic EUV wave train, not only appeared inside the CME bubble but also delayed the rapid expansion of the CME bubble by about 4 minutes. In addition, it was composed of multiple coherent wavefronts rather than a single pulse. These observational characteristics suggest that the observed large-scale quasi-periodic EUV wave train was unlikely excited by the associated CME. Since the observed EUV wave train has a similar period to the quasi-periodic pulsation of the associated flare, we more tend to regard that it was excited by the quasi-periodic pressure pulses launched by the intermittent energy releasing-process in the flare \citep{2008SoPh..253..215V}. If in that case, to the best of our knowledge, the present event should be a rare case showing strong evidence for supporting the flare-ignited EUV wave scenario. A possible detection of flare-ignited EUV wave event was presented by \cite{2008SoPh..253..305M} using radio observations, where the authors observed a type II radio burst which was well-timed with the associated flare but lagged the impulsive acceleration phase of the associated CME of about 10--20 minutes. Other ambiguous evidence were also documented as flare-ignited shocks, such as type II radio burst without any associated CME \citep{2012ApJ...746..152M}, pseudo CMEs \citep{2019SoPh..294...73E}, and null-point reconnection in fan-spine magnetic systems \citep{2016ApJ...828...28K}. Recent high spatiotemporal observations indicated that the generation of large-scale EUV waves does not need the presence of CMEs, they can be launched by coronal jets (\cite{,2018ApJ...861..105S}), sudden loop expansions due to the impingement of other eruptions \citep{2017ApJ...851..101S,2018ApJ...860L...8S,2018MNRAS.480L..63S}, or loop expansion of the reconnected open loop in coronal jets \citep{2015ApJ...804...88S}.

We note that large-scale quasi-periodic EUV wave trains propagating across a large fraction of the solar surface have only been reported in two studies \citep{2012ApJ...753...52L,2019ApJ...873...22S}. \cite{2012ApJ...753...52L} first observed an ambiguous example in the 2010 September 8--9 event and they proposed that the multiple wavefronts were within a broad global EUV wave propagating ahead of the associated CME, and they were driven by the strong lateral, downward expansion effect of the CME. For their scenario, it is hard to understand how can a single CME excite multiple large-scale EUV wavefronts. In addition, it is also difficult to understand the relationship between the multiple wavefronts and the broad global EUV wave. \cite{2019ApJ...873...22S} reported an unambiguous large-scale quasi-periodic EUV wave train on 2012 April 24, in which multiple coherent EUV wavefronts appeared consecutively at the periphery region of the active region and propagated simultaneously along a group of closed-loop system and on the solar surface. The authors found that the period of the wave train showed a large difference with the pulsations in the associated flare, but it was similar to the unwinding period of the helical thread of an erupting filament in the source region. Therefore, they proposed that the observed large-scale quasi-periodic EUV wave train was excited by the successive opening of the twisted threads of the eruption filament. Besides, the authors also proposed that such kind of EUV wave train should be different than the so-called QFP wave train due to several distinct observational characteristics that the latter do not have, including 1) the wave signal can be simultaneously observed at all the EUV channels of AIA, 2) the wavefronts propagate perpendicular to and parallel to coronal magnetic field lines, and 3) the intensity profiles and amplitudes of the wavefronts are more similar to typical large-scale EUV waves. Therefore, they called such kind of EUV wave as a large-scale quasi-periodic EUV wave train to show the difference between QFP wave trains.

The narrow quasi-periodic EUV wave train along the closed-loop system should be a typical flare-ignited QFP wave train as those have been documented in the literature \citep[see,][, and references therein]{2014SoPh..289.3233L}. Comparing with large-scale quasi-periodic EUV wave trains, the narrow quasi-periodic EUV wave train was only observed at the AIA 171 \AA\ channel with a small intensity amplitude of about 5\%, and propagated along the coronal loop. There is a main period in the narrow EUV wave train, which is of $113 \pm $\speed{8}. Obviously, the period of the narrow wave train was similar to the pulsation in the associate flare. This suggests that the narrow EUV wave train was possibly excited by the quasi-periodic pressure pulses launched by the intermittent energy-releasing process in the flare. It is noted that the appearance of the narrow quasi-periodic EUV wave train was slightly after the period signal detected in the flare, which indicates that some necessary time is needed for the wave signal to steep to a certain extent to be detected by {\em SDO}/AIA. A similar wave signal was also detected after the flare duration (see Figure 6 (c) and the arrows in Figure 6 (d)). It is unclear this was caused by another small flare or small energy release process in the late phase of the main flare.

\begin{acks}
We thank the anonymous referee for providing detailed suggestions that helped improve the paper. We also thank the {\em SDO} and {\em SOHO} teams for the data supply. This work is supported by the Yunnan Science Foundation (2017FB006), the National Key R\&D Program of China (2019YFA0405000), the Natural Science Foundation of China (12173083,11922307,11773068,11633008), the Yunnan Science Foundation for Distinguished Young Scholars (202101AV070004),  the Specialized Research Fund for State Key Laboratories, the Open Research Program of CAS Key Laboratory of Solar Activity (KLSA202017) and the West Light Foundation of Chinese Academy of Sciences.
\end{acks}


\end{article} 
\end{document}